\documentclass[twocolumn]{aastex631}

\newcommand{\teff}{$T_{\rm eff}$}

\newcommand{\logg}{$\log g$}

\newcommand{\feh}{$\rm{[Fe/H]}$}

\newcommand{\vmic}{$v_{t}$}

\newcommand{\qq}{$\mathrm{q}^{2}$}

\newcommand{\sm}{$\rm{M_{\odot}}$}

\newcommand{\vsini}{$v \sin i$}

\newcommand{\be}{\begin{equation}}
\newcommand{\ee}{\end{equation}}
\newcommand{\ben}{\begin{eqnarray}}
\newcommand{\een}{\end{eqnarray}}
\newcommand{\bfg}{\begin{figure}}
\newcommand{\efg}{\end{figure}}

\newcommand{\gaiaedr}{\textit{Gaia} EDR3}

\newcommand{\RHK}{$\log R^{'}_{\rm{HK}}$}

\newcommand{\li}{$A$(Li)}
\newcommand{\linlte}{$A$(Li)$_{\rm{NLTE}}$}
\newcommand{\vtd}{$\Delta v_{\rm{3D}}$}

\newcommand{\g}{$G$}

\received{July 28, 2021}
\revised{August 17, 2021}
\accepted{\today}
\shorttitle{Rocky planet engulfment in the binary system HIP 71726-37}
\shortauthors{Yana Galarza et al.}
\graphicspath{{./}{figures/}}
\begin{document}

\title{Evidence of rocky planet engulfment in the wide binary system HIP 71726/HIP 71737}

\correspondingauthor{J. Yana Galarza}
\email{ramstojh@alumni.usp.br}

\author[0000-0001-9261-8366]{Jhon Yana Galarza}
\affiliation{Universidade de S\~ao Paulo, Departamento de Astronomia do IAG/USP, \\
Rua do Mat\~ao 1226 Cidade Universit\'aria, 05508-900 S\~ao Paulo, SP, Brazil }
\affiliation{Center for Radio Astronomy and Astrophysics Mackenzie (CRAAM), \\ Mackenzie Presbyterian University, Rua da Consolação, 896, S\~ao Paulo, Brazil}

\author[0000-0002-7795-0018]{Ricardo L\'opez-Valdivia}
\affiliation{The University of Texas at Austin, Department of Astronomy, \\ 
2515 Speedway, Stop C1400, Austin, TX 78712-1205}

\author[0000-0002-4933-2239]{Jorge Mel\'endez}
\affiliation{Universidade de S\~ao Paulo, Departamento de Astronomia do IAG/USP, \\
Rua do Mat\~ao 1226 Cidade Universit\'aria, 05508-900 S\~ao Paulo, SP, Brazil }

\author[0000-0002-1387-2954]{Diego Lorenzo-Oliveira}
\affiliation{Universidade de S\~ao Paulo, Departamento de Astronomia do IAG/USP, \\
Rua do Mat\~ao 1226 Cidade Universit\'aria, 05508-900 S\~ao Paulo, SP, Brazil }



\begin{abstract}
Binary stars are supposed to be chemically homogeneous, as they are born from the same molecular cloud. However, high precision chemical abundances show that some binary systems display chemical differences between the components, which could be due to planet engulfment. In this work, we determine precise fundamental parameters and chemical abundances for the binary system HIP 71726/ HIP 71737. Our results show that the pair is truly conatal, coeval and comoving. We also find that the component HIP 71726 is more metal-rich than HIP 71737 in the refractory elements such as iron, with $\Delta$\feh\ $= 0.11 \pm 0.01$ dex. Moreover, HIP 71726 has a lithium abundance 1.03 dex higher than HIP 71737, which is the largest difference in Li detected in twin-star binary systems with $\Delta$\teff\ $\leq$ 50 K. The ingestion of $9.8^{+2.0}_{-1.6}$ M$_{\oplus}$ of rocky material fully explains both the enhancement in refractory elements and the high Li content observed in HIP 71726, thereby reinforcing the planet engulfment scenario in some binary systems. 
\end{abstract}



\section{Introduction} 
Chemical tagging is one of the main goals of large surveys such as GALAH \citep{DeSilva:2015MNRAS.449.2604D}, APOGEE \citep{Majewski:2017AJ....154...94M}, \textit{Gaia-ESO} \citep{Gilmore:2012Msngr.147...25G}, etc. As per this method, we could identify the birthplaces of stars in the Galaxy using simply their chemical composition. The technique is mainly based on two assumptions: (i) the chemically homogeneity of the proto-cloud from which stars form; (ii) the unique chemical signatures of formed stars compared to other star groups. In this context, both open clusters and wide binaries are ideal laboratories for testing these assumptions, since they are considered coeval and conatal stars \citep[e.g.,][]{De-Silva:2007AJ....133.1161D, Ting:2012MNRAS.427..882T, Ness:2018ApJ...853..198N, Andrews:2017MNRAS.472..675A, Hawkins:2020MNRAS.492.1164H}. 

Despite the large number of wide binary systems identified in the last decade \citep[e.g.,][]{Tokovinin:2014AJ....147...87T, Andrews:2017MNRAS.472..675A, El-Badry:2021MNRAS.tmp..394E}, only a few works have dedicated to studying their chemical homogeneity using spectra of high resolving power. In this sense,  \citet{Laws:2001ApJ...553..405L} and \citet{Gratton:2001A&A...377..123G} were pioneers in determining chemical abundances of wide binaries using the differential analysis and spectra of very high resolving power ($R = 58\ 000 - 160\ 000$) and signal-to-noise ratio (SNR = $100-400$). Other studies have focused on analyzing a few elements such as iron, vanadium, and lithium \citep{Desidera:2004A&A...420..683D, Desidera:2006A&A...454..581D, Martin:2002ApJ...579..437M}. The most of these wide binaries are chemically homogeneous, i.e., the metallicity difference between the components ($\Delta$\feh) is $\sim$ 0.0 dex, however other pairs have $\Delta$\feh\  $\sim$ 0.1 dex. 

A more comprehensive chemical analysis of a large sample of wide binaries was carried out by \citet{Andrews:2019ApJ...871...42A}, who used APOGEE ($R$ = 22 500) spectra to determine the abundance of 14 elements. They found that the chemical abundances of the systems are consistent to the level of 0.1 dex, however the precision decrease significantly to $\sim0.3$ dex for some elements like S (see their Figure 10). \citet{Hawkins:2020MNRAS.492.1164H}, using spectra of higher resolving power ($R = 60\ 000$), obtained a better precision and showed that the metallicity differences between the components of their sample of wide binaries is centered at \feh\ $\sim$ 0.0 dex with a scatter of 0.05 dex, and extreme values from about $-$0.1 to $+$0.1 dex. There is still a debate concerning whether wide binaries with large separations are still bound. According to \citet{Jiang:2010MNRAS.401..977J}, binary systems with separations $\ga 10^{5}$ AU are unbound, however \citet{Kamdar:2019ApJ...884..173K} demonstrated through simulations that comoving pairs with 3D differential velocities (\vtd) $<2$ km s$^{-1}$ are probably bound. This is supported by \citet{Nelson:2021arXiv210412883N}, who found that chemical abundances of comoving pairs with large separations are also consistent to the level of $\sim 0.1$ dex. More recently, \citet{Espinoza-Rojas:2021arXiv210501096E} showed that binary systems, even those with some metallicity difference, are useful to validate chemical clocks. All these results suggest that wide binaries can be excellent laboratories for testing the chemical-tagging hypotheses. However, it will only be possible if spectra of high resolving power and signal-to-noise ratio are employed in the differential analysis to achieve high precision chemical abundances.

When one of the components in a binary system is richer in refractory elements than its companion, the effect can be directly attributed to engulfment of planetary material \citep{Pinsonneault:2001ApJ...556L..59P}. This scenario was suggested for several binary systems with $\Delta$\feh $\geq 0.02$ dex such as HD 219542, 16 Cygni, XO-2, WASP-94, and HAT-P-4 \citep{Gratton:2001A&A...377..123G, Ramirez:2011ApJ...740...76R, Biazzo:2015A&A...583A.135B, Ramirez:2015ApJ...808...13R, Teske:2016ApJ...819...19T, Teske:2016AJ....152..167T, Saffe:2017A&A...604L...4S}. 
\begin{table*}
    \centering
	\caption{$^{(\star)}$Stellar parameters for the binary system HIP 71726-37 determined through the differential technique relative to the Sun. $^{(\surd)}$Stellar parameters for HIP 71737 and the Sun, both relative to the primary component HIP 71726. $^{(\dagger)}$Trigonometric surface gravity \citep[][see their equation 3]{Yana-Galarza:2021MNRAS.504.1873Y}.}
	\begin{scriptsize}
	\begin{tabular}{ccccccccc} 
		\hline
		\hline
		ID                    &  SNR  & \teff         & \logg             & \logg $^{(\dagger)}$ & \feh                  & \vmic              &  RV                   &     \vsini       \\ %  & \linlte           
		                      &       & (K)           & (dex)             & (dex)                & (dex)                 &  (km s$^{-1}$)     &  (km s$^{-1}$)        &   (km s$^{-1}$)  \\ %  &  (dex)            
        \hline                                                                                                                                                                              %
		HIP 71726$^{(\star)}$ &  340  & 5957 $\pm$ 9  & 4.260 $\pm$ 0.024 & 4.245 $\pm$ 0.022    &    0.164  $\pm$ 0.008 & 1.24   $\pm$ 0.02  &  $-$11.4 $\pm$ 0.2    & 0.17 $\pm$ 0.30  \\ %  &  2.92 $\pm$ 0.05  
		HIP 71737$^{(\star)}$ &  360  & 5934 $\pm$ 9  & 4.350 $\pm$ 0.025 & 4.305 $\pm$ 0.022    &    0.049  $\pm$ 0.008 & 1.21   $\pm$ 0.02  &  $-$11.0 $\pm$ 0.2    & 0.10 $\pm$ 0.20  \\ %  &  1.89 $\pm$ 0.05  
		%\hdashline                                                                                                                                                                         %
		HIP 71737$^{(\surd)}$ &  360  & 5934 $\pm$ 8  & 4.330 $\pm$ 0.023 &                      & $-$0.105  $\pm$ 0.007 & 1.19   $\pm$ 0.02  &                       &                  \\ %  &                   
		Sun$^{(\surd)}$       &  450  & 5781 $\pm$ 8  & 4.440 $\pm$ 0.023 &                      & $-$0.163  $\pm$ 0.008 & 1.00   $\pm$ 0.02  &                       &                  \\ %  &                   
		\hline
		\hline
	\end{tabular}
	\end{scriptsize}
	\label{tab:fundamental_parameters}
\end{table*}

\begin{table*}
	\caption{Adopted metallicities, ages, masses, and chromospheric activity indices estimated for the binary system HIP 71726-37. $^{(\dagger)}$Ages and masses determined using the lowest metallicity (\feh\ = 0.049 dex) of the pair. $^{(\star)}$Ages and masses calculated adopting the mean of the metallicity (\feh\ = 0.107 dex) of the pair.  $^{(\clubsuit)}$Ages computed from the age-activity relation of \citet{Diego:2016A&A...594L...3L}.}
	\centering
	\begin{scriptsize}
	\begin{tabular}{ccccccccc} 
		\hline
		\hline
		ID                    & \feh                & Age$_{\rm{Iso}}$       & Mass$_{\rm{Iso}}$      & $S_{\rm{MW}}$     &  \RHK             & Age$^{\clubsuit}_{\rm{HK}}$   \\
		                      & (dex)               & (Gyr)                  & (\sm)                  &                   &                   & (Gyr)                         \\
        \hline                                                                                                                                 
		HIP 71726             & $0.164 \pm 0.008$   & $4.20^{+0.26}_{-0.25}$ & $1.18^{+0.01}_{-0.01}$ & 0.147 $\pm$ 0.001 & -5.08 $\pm$ 0.07  &  $4.50^{+1.70}_{-1.30}$       \\
		HIP 71737             & $0.049 \pm 0.008$   & $5.10^{+0.28}_{-0.25}$ & $1.09^{+0.01}_{-0.01}$ & 0.156 $\pm$ 0.001 & -5.01 $\pm$ 0.07  &  $5.30^{+2.10}_{-1.40}$       \\
		HIP 71726$^{\dagger}$ & $0.049 \pm 0.008$   & $5.20^{+0.58}_{-0.50}$ & $1.14^{+0.01}_{-0.02}$ &                   &                   &  $5.30^{+2.00}_{-1.50}$       \\
		HIP 71726$^{\star}$   & $0.107 \pm 0.011$   & $4.60^{+0.45}_{-0.43}$ & $1.16^{+0.02}_{-0.01}$ &                   &                   &  $4.90^{+1.82}_{-1.39}$       \\
		HIP 71737$^{\star}$   & $0.107 \pm 0.011$   & $4.50^{+0.46}_{-0.47}$ & $1.12^{+0.01}_{-0.02}$ &                   &                   &  $4.90^{+1.78}_{-1.40}$       \\
		\hline
		\hline
	\end{tabular}
	\end{scriptsize}
	\label{tab:age_activity}
\end{table*}

The most of twin-star binaries analyzed with high resolving power ($R = \lambda/\Delta \lambda > 60\ 000$) have $\Delta$\feh\ up to $\sim 0.05$ dex. The system HIP 34407/HIP 34426 (\citealp{Ramirez:2019MNRAS.490.2448R}; see also \citealp{Nagar:2020ApJ...888L...9N, Espinoza-Rojas:2021arXiv210501096E}) and HD 240430/HD 240429 \citep{Oh:2018ApJ...854..138O} are the exception with $\Delta$\feh $\sim 0.2$ dex. The Fe-rich component in these systems is enhanced in refractory elements relative to its companion. Notably, the Li abundance (\li) in HD 240430 is larger than HD 240429 by $\sim0.5$ dex, likely due to a planet accretion of 15 M$_{\oplus}$ of rocky material. In particular, this scenario had already been studied by \citet{Gratton:2001A&A...377..123G}, who analyzed the system HD 219542A-B ($\Delta$\feh $= 0.09$ dex) and reported a very high enhancement in Li abundance ($\Delta$Li = 1.35 dex) in HD 219542A. 

Based on the results discussed above, the components of binary systems with $\Delta$\feh\ $\geq 0.02$ dex (or somewhat higher, depending on the achieved precision) can be affected by mechanisms that change their original composition, thereby questioning the viability of chemical tagging for these pairs. In this work, we report the discovery of abundance anomalies in the wide binary system HIP 71726/HIP 71737 (hereafter HIP 71726-37), whose components of G type stars show an unusual differential chemical abundance that suggest a rocky planet engulfment by the Fe-rich component (HIP 71726).

%%%%%%%%%%%%%%%%%%%%%%%%%%%%%%%%%%%%%%%%%%%
\section{Sample selection and observations}
We selected our pair from the sample of wide binaries of \citet{El-Badry:2021MNRAS.tmp..394E}, after applying the color constraints given in \citet[][Table 1]{Yana-Galarza:2021MNRAS.504.1873Y}, which were specially established for searching solar twins. We estimated the projected separation in $\sim$16764 AU using the \textsc{astropy} package, which employs coordinates and distances as input parameters. The coordinates and parallaxes are adopted from \gaiaedr\ \citep{GaiaEDR3:2021A&A...649A...1G}. The distance ($\sim$ 55 pc) is calculated inverting the parallax. We also determined the Galactic orbits for HIP 71726-37 using the \textsc{Gala}\footnote{\url{https://github.com/adrn/gala}} code \citep{gala}. The components of the pair have similar maximum vertical excursion ($z_{\rm{max}} = (0.205, 0.201)$ kpc), eccentricity ($0.137, 0.137$), perigalacticon (($10.200, 10.212$) kpc) and apogalacticon (($7.745, 7.748$) kpc) values. The Galactic space velocities are $U, V, W = (20.55, 20.52), (4.51, 4.69), (-20.60, -20.31)$ km s$^{-1}$, which places the wide binary within the thin-disk kinematic distribution on the Toomre diagram $V$ vs. $\sqrt{U^{2}+W^{2}}$ \citep[e.g.,][see their Figure 3]{Ramirez:2007A&A...465..271R}. The 3D velocity difference ($\Delta v_{\rm{3D}}$) is $0.34\pm0.07$ km s$^{-1}$, which takes into account the space velocities of each component. These results indicate that the binary system is bound and truly comoving, since the components also share identical radial velocities (see Table \ref{tab:fundamental_parameters}).

The spectra of our comoving pair were obtained using the Robert G. Tull Coud\'e Spectrograph \citep[][TS23]{Tull:1995PASP..107..251T} on the 2.7-m Harlam J. Smith Telescope at the McDonald Observatory during the months of February and March, 2021. The solar spectra were obtained through the reflected light from the Moon. We reduced the TS23 spectra employing our own scripts\footnote{\url{https://github.com/ramstojh/TS23-reduction}} based on \textsc{PyRAF}, which perform the standard procedure, i.e., creation of the master flat, flat-field correction, sky subtraction, order extraction, etc. The resulting spectra have high resolving power $R = 60\ 000$ with SNR = 450 for the Sun and SNR $\sim 350$ for each component of the binary system.

%%%%%%%%%%%%%%%%%%%%%%%%%%%%%%%%
\section{Fundamental parameters}
The radial/barycentric velocity correction and normalization of the TS23 spectra were carried out using our own semi-automatic \textsc{Python} scripts described in \citet{Yana-Galarza:2021MNRAS.504.1873Y}. The Equivalent Widths (EWs) were measured manually through Gaussian fits to the line profile using the \textsc{KAPTEYN} kmpfit Package \citep{KapteynPackage}, which is also implemented in our scripts. The technique is based on the line-by-line measurements choosing pseudo-continuum regions of 6 \AA\ in order to achieve a high precision. We adopted the line list of \citet{Melendez:2014ApJ...791...14M}, which includes information about the excitation potentials, oscillator strengths, laboratory $\log{gf}$ values, and hyperfine structure corrections \citep{Asplund:2009ARA&A..47..481A, McWilliam:1998AJ....115.1640M, Cohen:2003ApJ...588.1082C}. The adopted line list is composed of weak to moderate-strength lines (EW $<$ 130 m\AA).

We determined the effective temperature (\teff), surface gravity (\logg), metallicity (\feh), and microturbulence velocity (\vmic) using the automatic \qq\ \citep{Ramirez:2014A&A...572A..48R} code adopting the spectroscopic equilibrium technique. The \qq\ was configured to employ the Kurucz \textsc{ODFNEW} model atmospheres \citep{Castelli:2003IAUS..210P.A20C} and the 2019 local thermodynamic equilibrium (LTE) code \textsc{moog} \citep{Sneden:1973PhDT.......180S}. The results are summarized in Table \ref{tab:fundamental_parameters}, where the first two rows are the stellar parameters of the pair relative to the Sun and the last two are the stellar parameters of HIP 71737 and the Sun relative to the primary component (HIP 71726). Both differential methods result in identical stellar parameters for HIP 71737, and in the case of the Sun, we fully recover the solar stellar parameters (\teff\ = 5777 K, \logg\ = 4.44 dex, \feh\ = 0.0 dex, \vmic\ = 1.0 km s$^{-1}$).
\begin{table*}
    \centering
	\caption{$^{(\star)}$Differential chemical abundances relative to the Sun. $^{(\surd)}$Elemental abundance differences between HIP 717237 and HIP 717226. The las two columns are the predicted chemical abundance from a model of planet engulfment and the dust condensation temperature of elements, respectively. $^{(\dagger)}$Non-LTE absolute abundances of lithium (\linlte).}
	\begin{scriptsize}
	\begin{tabular}{cccccccc} 
		\hline
		\hline
		Element           &   Number    &   Z   &  HIP 717226$^{(\star)}$  &  HIP 717237$^{(\star)}$  & HIP 717237-27$^{(\surd)}$  &     Model           &  $T_{\rm{Cond}}$  \\   
		                  &  of lines   &       &      $\Delta$[X/H] (dex) &   $\Delta$[X/H] (dex)    &    $\Delta$[X/H] (dex)     & $\Delta$[X/H] (dex) &       (K)         \\   
        \hline                                                                                                                                                                                                                                                    
		Li$^{(\dagger)}$  &      1      &   3   &    $2.920 \pm 0.060$     &     $1.890 \pm 0.060$    &   $ 1.030 \pm 0.085$       &      $-0.970$       &        1142       \\ 
		C                 &      3      &   6   &    $0.035 \pm 0.011$     &     $0.070 \pm 0.009$    &   $ 0.030 \pm 0.010$       &      $-0.006$       &        40         \\   
		O                 &      3      &   8   &    $0.068 \pm 0.014$     &     $0.043 \pm 0.016$    &   $-0.026 \pm 0.015$       &      $-0.057$       &        180        \\  
		Na                &      3      &   11  &    $0.149 \pm 0.019$     &     $0.103 \pm 0.019$    &   $-0.042 \pm 0.009$       &      $-0.097$       &        958        \\  
		Mg                &      3      &   12  &    $0.228 \pm 0.025$     &     $0.088 \pm 0.010$    &   $-0.132 \pm 0.009$       &      $-0.150$       &        1336       \\ 
		Al                &      3      &   13  &    $0.232 \pm 0.016$     &     $0.096 \pm 0.009$    &   $-0.132 \pm 0.008$       &      $-0.171$       &        1653       \\ 
		Si                &     12      &   14  &    $0.195 \pm 0.007$     &     $0.089 \pm 0.006$    &   $-0.105 \pm 0.005$       &      $-0.155$       &        1310       \\ 
		S                 &      3      &   16  &    $0.110 \pm 0.010$     &     $0.055 \pm 0.031$    &   $-0.042 \pm 0.010$       &      $-0.065$       &        664        \\  
		K                 &      1      &   19  &    $0.180 \pm 0.014$     &     $0.104 \pm 0.011$    &   $-0.088 \pm 0.013$       &      $-0.080$       &        1006       \\ 
		Ca                &     10      &   20  &    $0.181 \pm 0.012$     &     $0.057 \pm 0.012$    &   $-0.114 \pm 0.008$       &      $-0.168$       &        1517       \\ 
		Sc                &      8      &   21  &    $0.273 \pm 0.006$     &     $0.100 \pm 0.008$    &   $-0.168 \pm 0.009$       &      $-0.142$       &        1659       \\ 
		Ti                &     21      &   22  &    $0.209 \pm 0.012$     &     $0.086 \pm 0.009$    &   $-0.125 \pm 0.009$       &      $-0.143$       &        1582       \\ 
		VI                &      5      &   23  &    $0.203 \pm 0.017$     &     $0.083 \pm 0.013$    &   $-0.116 \pm 0.018$       &      $-0.189$       &        1429       \\ 
		Cr                &     18      &   24  &    $0.175 \pm 0.011$     &     $0.062 \pm 0.011$    &   $-0.107 \pm 0.009$       &      $-0.156$       &        1296       \\ 
		Mn                &      7      &   25  &    $0.120 \pm 0.012$     &     $0.056 \pm 0.013$    &   $-0.072 \pm 0.010$       &      $-0.117$       &        1158       \\ 
		Fe                &    100      &   26  &    $0.164 \pm 0.008$     &     $0.049 \pm 0.008$    &   $-0.105 \pm 0.007$       &      $-0.145$       &        1334       \\ 
		Co                &      7      &   27  &    $0.188 \pm 0.010$     &     $0.083 \pm 0.012$    &   $-0.115 \pm 0.012$       &      $-0.128$       &        1352       \\ 
		Ni                &     15      &   28  &    $0.193 \pm 0.008$     &     $0.065 \pm 0.008$    &   $-0.124 \pm 0.008$       &      $-0.145$       &        1353       \\ 
		Cu                &      2      &   29  &    $0.119 \pm 0.021$     &     $0.042 \pm 0.020$    &   $-0.074 \pm 0.009$       &      $-0.110$       &        1037       \\ 
		Zn                &      2      &   30  &    $0.151 \pm 0.020$     &     $0.070 \pm 0.020$    &   $-0.078 \pm 0.025$       &      $-0.065$       &        726        \\  
		Sr                &      1      &   38  &    $0.165 \pm 0.019$     &     $0.009 \pm 0.019$    &   $-0.165 \pm 0.017$       &      $-0.181$       &        1464       \\ 
		Y                 &      2      &   39  &    $0.144 \pm 0.027$     &     $0.021 \pm 0.032$    &   $-0.114 \pm 0.016$       &      $-0.148$       &        1659       \\ 
		Ba                &      3      &   56  &    $0.196 \pm 0.023$     &     $0.057 \pm 0.020$    &   $-0.130 \pm 0.012$       &      $-0.140$       &        1455       \\ 
		La                &      2      &   57  &    $0.350 \pm 0.035$     &     $0.244 \pm 0.033$    &   $-0.114 \pm 0.035$       &      $-0.185$       &        1578       \\ 
		Ce                &      4      &   58  &    $0.118 \pm 0.020$     &     $0.015 \pm 0.020$    &   $-0.106 \pm 0.010$       &      $-0.168$       &        1478       \\ 
		Nd                &      2      &   60  &    $0.190 \pm 0.011$     &     $0.077 \pm 0.011$    &   $-0.107 \pm 0.014$       &      $-0.175$       &        1602       \\ 
		Sm                &      3      &   62  &    $0.268 \pm 0.020$     &     $0.185 \pm 0.019$    &   $-0.086 \pm 0.016$       &      $-0.162$       &        1590       \\ 
		Eu                &      1      &   63  &    $0.229 \pm 0.019$     &     $0.017 \pm 0.018$    &   $-0.207 \pm 0.035$       &      $-0.156$       &        1356       \\ 
		\hline
		\hline
	\end{tabular}
	\end{scriptsize}
	\label{tab:chemical_abundances}
\end{table*}

We also computed the ages and masses employing the Yonsei–Yale isochrones of stellar evolution \citep{Yi:2001ApJS..136..417Y, Demarque:2004ApJS..155..667D}, which is already implemented in the \qq\ code. Unlike the traditional isochronal age calculations that use only absolute $V$ magnitude or spectroscopic \logg\ as input parameter, we use both parameters in the \qq\ code. The parallaxes are adopted from \gaiaedr\ and we corrected them by subtracting $-15 \pm 18\ \mu$as as suggested by \citealp{Stassun:2021ApJ...907L..33S}.

Additionally, we estimated the activity indices ($S_{\rm{HK, TS23}}$) measuring the flux in the \ion{Ca}{2} H\&K lines following the prescription given in \citet{Wright:2004ApJS..152..261W}. The $S_{\rm{HK, TS23}}$ was converted to the Mount Wilson system ($S_{\rm{MW}}$) using the equation (4) of \citet{Yana-Galarza:2021MNRAS.504.1873Y}. Then, we determined the chromospheric indices (\RHK) following the procedure of \citet{Noyes:1984ApJ...279..763N}. Finally, the chromospheric ages are estimated from the age-mass-metallicity-activity correlation derived by \citet[][see their equation 1]{Diego:2016A&A...594L...3L}, which takes mass and metallicity biases into account. The results are shown in Table \ref{tab:age_activity}.

Both the isochronal and chromospheric ages indicate a difference of $\sim 1$ Gyr between the components, which could signify that the system is not coeval. However, if the Fe-rich component (HIP 71726) was enhanced with refractory elements, its iron content does not reflect the metallicity of the entire star, but only that of the convective zone, which artificially changes the age because an overestimated metallicity is being used to obtain isochronal and chromospheric ages. Therefore, to assess the coevality of our system, we performed a simple experiment based on estimating ages and masses using the mean and the lowest \feh\ of the pair, assuming that these are the `real' metallicities. As shown in Table \ref{tab:age_activity}, the isocronal ages of the components are identical between them in both cases, with a difference of only 0.1 Gyr. Similar results are found for the chromospheric ages. Therefore, we can consider that our binary system is truly coeval.

%%%%%%%%%%%%%%%%%%%%%%%%%%%%%%
\section{Chemical composition}
\begin{figure}
 \includegraphics[width=1.01\columnwidth]{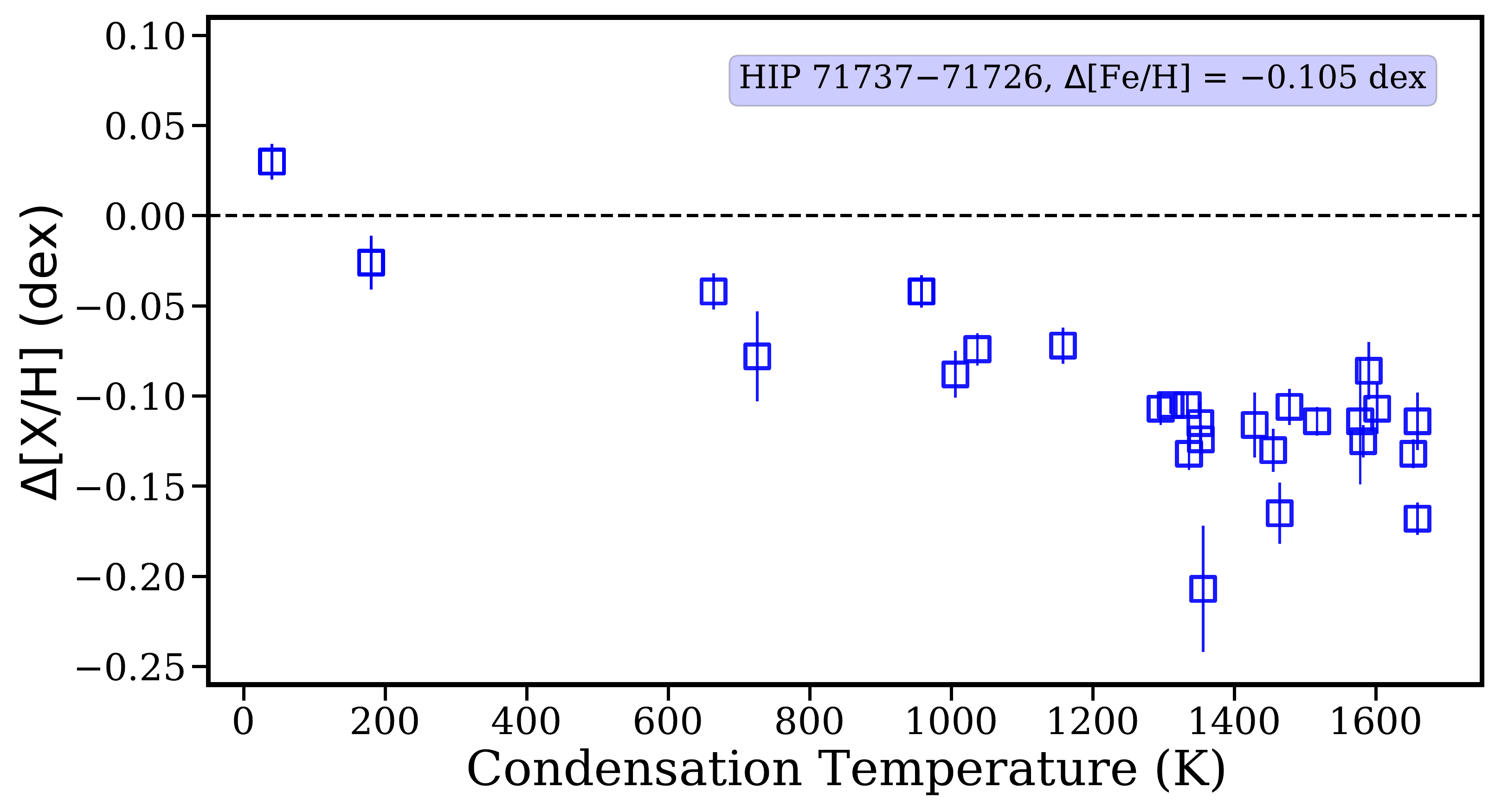}
 \centering
 \caption{Differential chemical abundances of HIP 71737 relative to HIP 71726 as a function of dust condensation temperature \citep{Lodders:2003ApJ...591.1220L}.}
 \label{fig:Delta_chemical}
\end{figure}

The chemical abundance for 26 elements (C, O, Na, Mg, Al, Si, S, K, Ca, Sc, Ti, V, Cr, Mn, Co, Ni, Cu, Zn, Sr, Y II, Ba II, La II, Ce II, Nd II, Sm II and Eu II) were estimated using the \qq\ code, which was configured to perform corrections of NLTE for the O I triplet \citep{Ramirez:2007A&A...465..271R} and take into account the hyperfine structure for V, Mn, Co, Cu, Y II, Ba II, La II and Eu II \citep{Melendez:2014ApJ...791...14M}. The internal precision for most differential abundances is $\sim 0.01$ dex, which was calculated following the prescription given in \citet{Epstein:2010ApJ...709..447E}. It is important to highlight that there is only one spectral line available for the elements K I, Sr I, and Eu II, therefore their abundances were estimated assuming various pseudo-continuum levels. The adopted abundance for each of these three elements is the average of all the measurements while the observational uncertainty is the standard deviation. Special attention was paid to regions where telluric lines can affect the measurements (e.g., K I  at 7698.97 \AA). The differential chemical abundances for the components of the pair relative to the Sun are summarized in Table \ref{tab:chemical_abundances}. We also estimated the elemental abundance differences between HIP
717237 and HIP 717226 in order to improve the precision. This process was performed adopting the HIP 717226 as reference. The result is shown in Figure \ref{fig:Delta_chemical}, where it can be clearly seen that the differential abundance pattern of HIP 71737 is deficient in refractory elements (with high condensation temperature) relative to HIP 71726.

\begin{figure}
 \includegraphics[width=0.9\columnwidth]{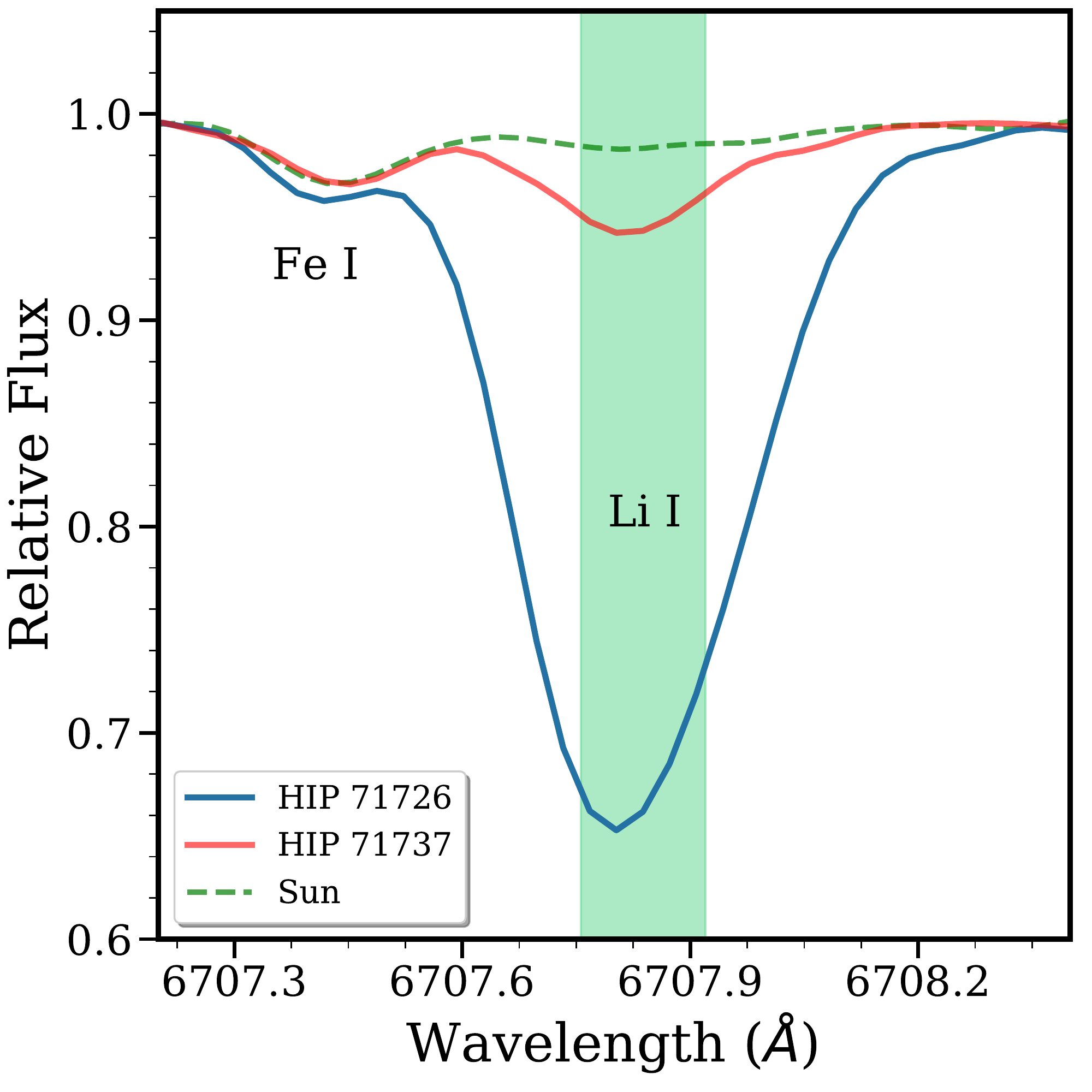}
 \centering
 \caption{Observed spectra of HIP 71726 (blue line) and HIP 71737 (red line) around the lithium line (shaded region). For comparison, the Sun’s spectrum is plotted as a green dashed line.}
 \label{fig:Li_binaries}
\end{figure}

\citet{Carlos:2016A&A...587A.100C, Carlos:2019MNRAS.tmp..667C} showed that the surface Li abundance (\li) in solar twins decreases with age, but with some scatter, probably due to somewhat different stellar masses and metallicities. We determined the \li\ for the binary system strictly following the procedure given in \citet{Yana_Galarza:2016A&A...589A..17Y} and \citet{Carlos:2016A&A...587A.100C}, i.e., synthetizing the region around the asymmetric 6707.75 \AA\ Li line using the LTE \textsc{moog} code and the Kurucz model atmosphere. The non-LTE (NLTE) lithium abundance (\linlte) was obtained using the INSPECT database\footnote{\url{www.inspect-stars.com}} based on corrections of \citet{Lind:2009A&A...503..541L}. The \linlte\ values are shown in Table \ref{tab:chemical_abundances} while in Figure \ref{fig:Li_binaries} are depicted the Li lines in the observed spectra of HIP 71726-37. Conspicuously, the Fe-rich star has \li\ 1.03 dex higher than HIP 71737. To our knowledge, this is the largest difference in \li\ observed in twin-star wide binaries with $\Delta$\teff\ $\leq 50$ K. The binary HD 219542A-B analyzed by \cite{Gratton:2001A&A...377..123G} has a larger difference in Li (1.35 dex), but it has $\Delta$\teff\ $\sim 300$ K.

It is very well known that the \li\ increases with \teff\ \cite[e.g.,][]{Ramirez:2012ApJ...756...46R, Bragaglia:2018AA...619A.176B}, and also its depletion is higher for stars with larger metallicities (\citealp{Carlos:2019MNRAS.tmp..667C}, \citealp{Castro:2009A&A...494..663C}, \citealp{Ramirez:2012ApJ...756...46R}, see their Figure 5 for thin disk stars with \feh\ $>$ 0). Based on these evidences, the Fe-rich star HIP 71726 should have depleted more Li than its companion, however this is not the case. According to the observational trend of Li with mass by \cite{Carlos:2019MNRAS.tmp..667C}, the different masses (0.09 \sm) of the pair suggest a difference of $0.3 \pm 0.1$ dex in \li\ \citep[see also][]{Ramirez:2012ApJ...756...46R, Delgado_Mena:2015A&A...576A..69D, Castro:2016A&A...590A..94C}, which only explains in part the observed Li in HIP 71726. Another possibility is different amounts of rotational mixing between the components. Nonetheless, after fitting the line profile of five iron lines: 6027.050 \AA, 6093.644 \AA, 6151.618 \AA, 6165.360 \AA, 6705.102\AA\ (see \citealp{Leo:2016A&A...592A.156D} and \citealp{Yana_Galarza:2016A&A...589A..17Y}), we found similar projected \vsini\ velocities for the components (see Table \ref{tab:fundamental_parameters}). The only plausible scenario that remains to explain the Li difference is the ingestion of rocky material by the Fe-rich star.

%%%%%%%%%%%%%%%%%%%%%%%%%%%
\section{Planet engulfment}
The abundance patterns (versus condensation temperature) found by \citet{Melendez:2009ApJ...704L..66M, Melendez:2017A&A...597A..34M, Oh:2018ApJ...854..138O, Tucci:2019A&A...628A.126M, Ramirez:2019MNRAS.490.2448R} and \citet{Saffe:2015A&A...582A..17S} in solar twins and twin-star binary systems have been linked to the formation of planets (for an alternative explanation regarding dust cleansing, see \citealp{Gustafsson:2018A&A...620A..53G, Gustafsson:2018A&A...616A..91G}). In particular, it is argued that the formation of rocky planets removes refractory elements from the hosted star. Thus, in a binary system, the component with more rocky planets is likely to be deficient in refractory elements. Based on this hypothesis, the deficiency of refractory elements of HIP 71737 would result from host rocky planets (see Figure \ref{fig:Delta_chemical}), although no planets have been detected yet in this system.
\begin{figure*}
 \includegraphics[width=1.8\columnwidth]{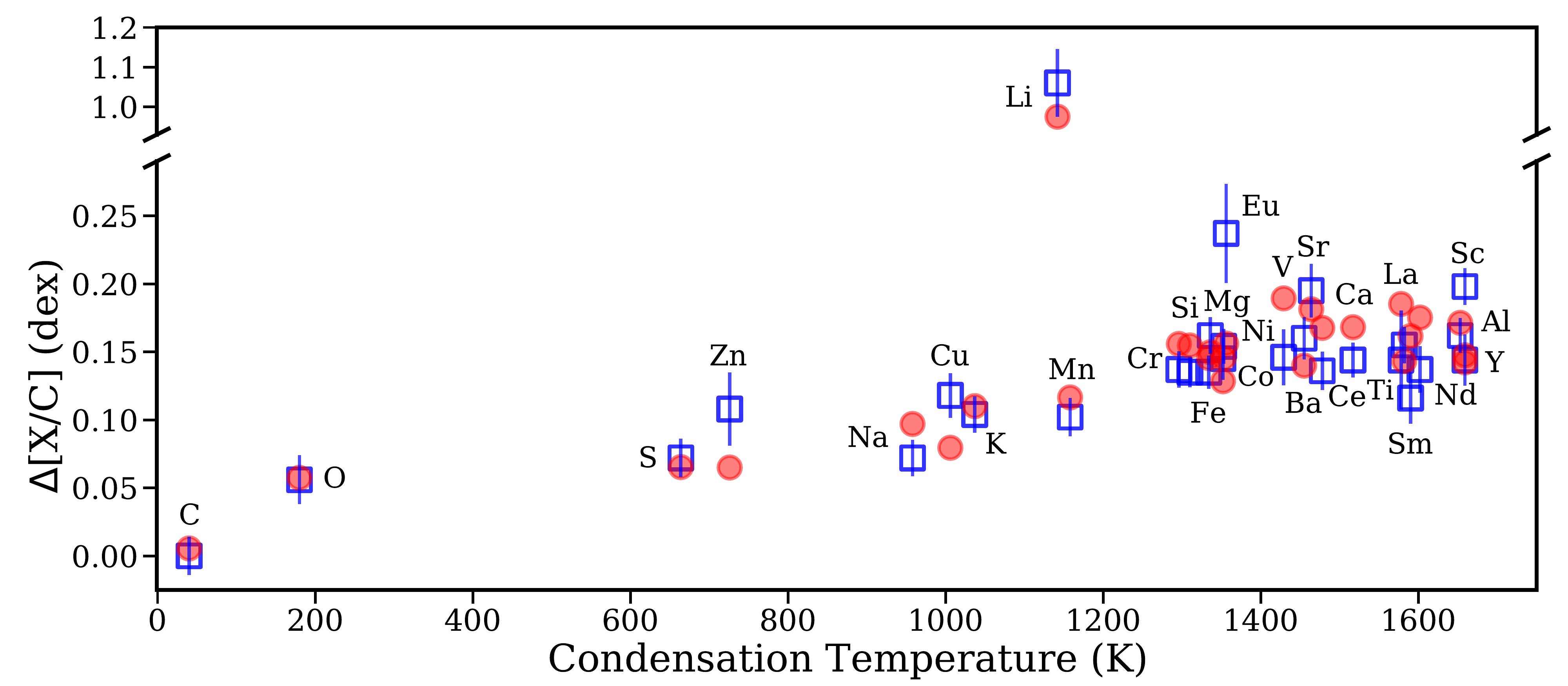}
 \centering
 \caption{Comparison of the abundances of HIP 71726 relative to HIP 71737 (blue squares) as a function of dust condensation temperature, and the predicted abundances (red circles) estimated from a planetary engulfment of $9.8^{+2.0}_{-1.6}$ M$_{\oplus}$.}
 \label{fig:XC_TC_binaries}
\end{figure*}

An alternative to the metal depletion scenario is a rocky planet engulfment, where the Fe-rich star HIP 71726 would have ingested planetary material rich in refractory elements. To validate this scenario, we used our code \textsc{terra}\footnote{\url{https://github.com/ramstojh/terra}} \citep{Yana_Galarza:2016A&A...589A..65G}, which is updated with the isochrones of stellar evolution of \citet{Yi:2001ApJS..136..417Y} and the more recent solar abundances of \citet{Asplund:2021arXiv210501661A}. In brief, the methodology consist in first determining the convective mass of a star through interpolations in metallicity and mass. Then, the code estimates the amount of rocky material necessary to reproduce the observed abundance pattern by adding meteoritic \citep{Wasson:1988RSPTA.325..535W} and terrestrial \citep{Allegre:2001E&PSL.185...49A} abundances patterns into its convective zone \citep[for more details see the appendix of][]{Yana_Galarza:2016A&A...589A..65G}.

The observed differential abundance pattern of HIP 71726 can be explained by a rocky planet engulfment of $9.8^{+2.0}_{-1.6}$ M$_{\oplus}$, which is a mixture of 7.0 M$_{\oplus}$ of chondrite-like composition and 2.8 M$_{\oplus}$ of Earth-like composition. As can be seen in Figure \ref{fig:XC_TC_binaries}, both the observed (open squares) and predicted (red circles) differential abundances of HIP 71726 are in good agreement.

The \li\ is also fully predicted by the planetary engulfment (see Figure \ref{fig:XC_TC_binaries}), however special attention must be paid to this point, since determining the total depleted Li abundance is a challenging task because we do not have information about the initial Li content at the time of accretion or the efficiency at which Li is depleted. Regardless, we carried out a simple experiment where it is assumed that the pair had experienced the same lithium depletion, i.e., they should have the same Li content (1.89 dex). The \li\ of HIP 71726 is 1.03 dex ([Li/C] = 1.06 dex) higher than HIP 71737 and an accretion of $9.8$ M$_{\oplus}$ result in 0.98 dex ([Li/C] = 0.97 dex), very consistent with the difference in Li observed. Similar analyses were performed by \citet{Melendez:2017A&A...597A..34M} and \citet{Oh:2018ApJ...854..138O} in the solar twin HIP 68468 and HD 240429/HD 240430, respectively. Albeit, these studies use the solar \li\ as reference.

If the planet engulfment scenario discussed above is correct, HIP 71726's rotation would be slightly faster \citep{Oetjens:2020A&A...643A..34O}. However, there are no significant differences in the \vsini\ of the components (see Table \ref{tab:fundamental_parameters}). This result does not invalidated our scenario because the planetary engulfment could happen when the star was young and its rotation was high, thus the small increase in angular momentum added by the planet has already been lost. Alternatively, the mass of the planet is too small to cause a significant increase in the star's angular momentum. In addition, according to \citet{Leo:2016A&A...592A.156D}, a tipical \vsini\ for a star with $\sim$4.9 Gyr is $\sim$1.8 km s$^{-1}$ (e.g. HIP 1954 therein). Therefore, the very low values of the \vsini\ of the components (0.17 km s$^{-1}$ and 0.10 km s$^{-1}$) could be only a projection effect ($\sin{i} \sim 0$). This means that the components could have different rotational velocities, but due to their inclinations, the difference cannot be detected. 

%Further studies with higher resolving power than the TS23 are required to better distinguish if there is any significant difference in the rotational velocities of the pair.

%%%%%%%%%%%%%%%%%%%%%
\section{Summary and Conclusions}
The precise stellar parameters and chemical abundances for the binary system HIP 71726-37 were estimated through a differential analysis relative to the Sun. To increase our precision, we also determined the stellar parameters of HIP 71737 and the Sun relative to the primary component (HIP 71726). %The obtained results in both analysis are identical (see Table \ref{tab:fundamental_parameters}).

The binary system has $\Delta$\feh\ $\sim 0.1$ dex, which means that the pair is chemical inhomogeneous. We estimated the Galactic orbits for the system and found identical eccentricity, perigalacticon and apogalacticon, and maximum excursion vertical values for both components. The 3D velocity difference is 0.96 km s$^{-1}$, suggesting that the pair is truly comoving. The ages and masses were calculated using isochrones and the age-activity relation of \citet{Diego:2016A&A...594L...3L}. The slightly different ages ($\sim$1 Gyr) of the components seemed to suggest that they are not coeval. However, when we use the mean or the lower \feh\ of the pair, a identical age is obtained (see Table \ref{tab:age_activity}). This suggests that HIP 71726 may have been contaminated by some material rich in refractory elements. Therefore, based on the above results, we conclude that the binary system HIP 71726-37 is truly conatal, coeval and comoving.

The chemical abundance pattern (versus condensation temperature) of HIP 71737 is deficient in refractory elements relative to the Fe-rich HIP 71726 (see Figure \ref{fig:Delta_chemical}). Following \citet{Melendez:2009ApJ...704L..66M}, HIP 71737 could have formed rocky exoplanets. However, no planets have been detected in this system yet. An alternative explanation is a planet engulfment scenario, where the Fe-rich star would have ingested a planet rich in refractory elements. Indeed, a planetary engulfment of $9.8^{+2.0}_{-1.6}$ M$_{\oplus}$ reproduces well the differential chemical abundance observed in HIP 71726 and also explains its high lithium abundance. Our study confirms that high precision chemical abundances in binary systems is a powerful tool to constrain past dramatic events of planet accretion.

\begin{acknowledgments}
\textit{Research funding agencies:} J.Y.G. acknowledges the support from CNPq (142084/2017-4; 166042/2020-0). J.M. and D.L.O. thank the support from FAPESP (2018/04055-8; 2016/20667-8).
\end{acknowledgments}


\vspace{5mm}
\facility{McDonald Observatory, Harlam J. Smith 2.7-meter Telescope, TS23 spectrograph.}


\software{\textsc{numpy} \citep{van_der_Walt:2011CSE....13b..22V}, \textsc{matplotlib} \citep{Hunter:4160265}, \textsc{pandas} \citep{mckinney-proc-scipy-2010}, \textsc{astroquery} \citep{Ginsburg:2019AJ....157...98G}, \textsc{iraf} \citep{Tody:1986SPIE..627..733T}, \textsc{iSpec} \citep{Blanco:2014A&A...569A.111B, Blanco:2019MNRAS.486.2075B}, \textsc{Kapteyn} Package \citep{KapteynPackage}, \textsc{moog} \citep{Sneden:1973PhDT.......180S}, q$^{\textsc{2}}$ \citep{Ramirez:2014A&A...572A..48R}, \textsc{Gala} \citep{gala}}, \textsc{terra} \citep{Yana_Galarza:2016A&A...589A..65G}. \\%, \textsc{CERES} \citep{CERES:2017PASP..129c4002B}, \textsc{SP\_ACE} \citep{SPACE:2016A&A...587A...2B}, \textsc{isochrones} \citep{Morton:2015ascl.soft03010M}, \textsc{MESA} \citep{Dotter:2016ApJS..222....8D, Choi:2016ApJ...823..102C, Paxton:2011ApJS..192....3P, Paxton:2013ApJS..208....4P, Paxton:2015ApJS..220...15P}, \textsc{MultiNest} \citep{Feroz:2008MNRAS.384..449F, Feroz:2009CQGra..26u5003F, Feroz:2019OJAp....2E..10F}.}



%\appendix

%\section{Appendix information}


\bibliography{references}{}

\begin{thebibliography}{}
\expandafter\ifx\csname natexlab\endcsname\relax\def\natexlab#1{#1}\fi
\providecommand{\url}[1]{\href{#1}{#1}}
\providecommand{\dodoi}[1]{doi:~\href{http://doi.org/#1}{\nolinkurl{#1}}}
\providecommand{\doeprint}[1]{\href{http://ascl.net/#1}{\nolinkurl{http://ascl.net/#1}}}
\providecommand{\doarXiv}[1]{\href{https://arxiv.org/abs/#1}{\nolinkurl{https://arxiv.org/abs/#1}}}

\bibitem[{{All{\`e}gre} {et~al.}(2001){All{\`e}gre}, {Manh{\`e}s}, \&
  {Lewin}}]{Allegre:2001E&PSL.185...49A}
{All{\`e}gre}, C., {Manh{\`e}s}, G., \& {Lewin}, {\'E}. 2001, Earth and
  Planetary Science Letters, 185, 49, \dodoi{10.1016/S0012-821X(00)00359-9}

\bibitem[{{Andrews} {et~al.}(2019){Andrews}, {Anguiano}, {Chanam{\'e}},
  {Ag{\"u}eros}, {Lewis}, {Hayes}, \& {Majewski}}]{Andrews:2019ApJ...871...42A}
{Andrews}, J.~J., {Anguiano}, B., {Chanam{\'e}}, J., {et~al.} 2019, \apj, 871,
  42, \dodoi{10.3847/1538-4357/aaf502}

\bibitem[{{Andrews} {et~al.}(2017){Andrews}, {Chanam{\'e}}, \&
  {Ag{\"u}eros}}]{Andrews:2017MNRAS.472..675A}
{Andrews}, J.~J., {Chanam{\'e}}, J., \& {Ag{\"u}eros}, M.~A. 2017, \mnras, 472,
  675, \dodoi{10.1093/mnras/stx2000}

\bibitem[{{Asplund} {et~al.}(2021){Asplund}, {Amarsi}, \&
  {Grevesse}}]{Asplund:2021arXiv210501661A}
{Asplund}, M., {Amarsi}, A.~M., \& {Grevesse}, N. 2021, arXiv e-prints,
  arXiv:2105.01661.
\newblock \doarXiv{2105.01661}

\bibitem[{{Asplund} {et~al.}(2009){Asplund}, {Grevesse}, {Sauval}, \&
  {Scott}}]{Asplund:2009ARA&A..47..481A}
{Asplund}, M., {Grevesse}, N., {Sauval}, A.~J., \& {Scott}, P. 2009, \araa, 47,
  481, \dodoi{10.1146/annurev.astro.46.060407.145222}

\bibitem[{{Biazzo} {et~al.}(2015){Biazzo}, {Gratton}, {Desidera}, {Lucatello},
  {Sozzetti}, {Bonomo}, {Damasso}, {Gand olfi}, {Affer}, {Boccato}, {Borsa},
  {Claudi}, {Cosentino}, {Covino}, {Knapic}, {Lanza}, {Maldonado}, {Marzari},
  {Micela}, {Molaro}, {Pagano}, {Pedani}, {Pillitteri}, {Piotto}, {Poretti},
  {Rainer}, {Santos}, {Scandariato}, \& {Zanmar
  Sanchez}}]{Biazzo:2015A&A...583A.135B}
{Biazzo}, K., {Gratton}, R., {Desidera}, S., {et~al.} 2015, \aap, 583, A135,
  \dodoi{10.1051/0004-6361/201526375}

\bibitem[{{Blanco-Cuaresma}(2019)}]{Blanco:2019MNRAS.486.2075B}
{Blanco-Cuaresma}, S. 2019, \mnras, 486, 2075, \dodoi{10.1093/mnras/stz549}

\bibitem[{{Blanco-Cuaresma} {et~al.}(2014){Blanco-Cuaresma}, {Soubiran},
  {Heiter}, \& {Jofr{\'e}}}]{Blanco:2014A&A...569A.111B}
{Blanco-Cuaresma}, S., {Soubiran}, C., {Heiter}, U., \& {Jofr{\'e}}, P. 2014,
  \aap, 569, A111, \dodoi{10.1051/0004-6361/201423945}

\bibitem[{{Bragaglia} {et~al.}(2018){Bragaglia}, {Fu}, {Mucciarelli},
  {Andreuzzi}, \& {Donati}}]{Bragaglia:2018AA...619A.176B}
{Bragaglia}, A., {Fu}, X., {Mucciarelli}, A., {Andreuzzi}, G., \& {Donati}, P.
  2018, \aap, 619, A176, \dodoi{10.1051/0004-6361/201833888}

\bibitem[{{Carlos} {et~al.}(2016){Carlos}, {Nissen}, \&
  {Mel{\'e}ndez}}]{Carlos:2016A&A...587A.100C}
{Carlos}, M., {Nissen}, P.~E., \& {Mel{\'e}ndez}, J. 2016, \aap, 587, A100,
  \dodoi{10.1051/0004-6361/201527478}

\bibitem[{{Carlos} {et~al.}(2019){Carlos}, {Mel{\'e}ndez}, {Spina}, {dos
  Santos}, {Bedell}, {Ramirez}, {Asplund}, {Bean}, {Yong}, {Yana Galarza}, \&
  {Alves-Brito}}]{Carlos:2019MNRAS.tmp..667C}
{Carlos}, M., {Mel{\'e}ndez}, J., {Spina}, L., {et~al.} 2019, \mnras, 485,
  4052, \dodoi{10.1093/mnras/stz681}

\bibitem[{{Castelli} \& {Kurucz}(2003)}]{Castelli:2003IAUS..210P.A20C}
{Castelli}, F., \& {Kurucz}, R.~L. 2003, in IAU Symposium, Vol. 210, Modelling
  of Stellar Atmospheres, ed. N.~{Piskunov}, W.~W. {Weiss}, \& D.~F. {Gray},
  A20

\bibitem[{{Castro} {et~al.}(2016){Castro}, {Duarte}, {Pace}, \& {do
  Nascimento}}]{Castro:2016A&A...590A..94C}
{Castro}, M., {Duarte}, T., {Pace}, G., \& {do Nascimento}, J.~D. 2016, \aap,
  590, A94, \dodoi{10.1051/0004-6361/201527583}

\bibitem[{{Castro} {et~al.}(2009){Castro}, {Vauclair}, {Richard}, \&
  {Santos}}]{Castro:2009A&A...494..663C}
{Castro}, M., {Vauclair}, S., {Richard}, O., \& {Santos}, N.~C. 2009, \aap,
  494, 663, \dodoi{10.1051/0004-6361:20078928}

\bibitem[{{Cohen} {et~al.}(2003){Cohen}, {Christlieb}, {Qian}, \&
  {Wasserburg}}]{Cohen:2003ApJ...588.1082C}
{Cohen}, J.~G., {Christlieb}, N., {Qian}, Y.~Z., \& {Wasserburg}, G.~J. 2003,
  \apj, 588, 1082, \dodoi{10.1086/374269}

\bibitem[{{De Silva} {et~al.}(2007){De Silva}, {Freeman}, {Asplund},
  {Bland-Hawthorn}, {Bessell}, \& {Collet}}]{De-Silva:2007AJ....133.1161D}
{De Silva}, G.~M., {Freeman}, K.~C., {Asplund}, M., {et~al.} 2007, \aj, 133,
  1161, \dodoi{10.1086/511182}

\bibitem[{{De Silva} {et~al.}(2015){De Silva}, {Freeman}, {Bland-Hawthorn},
  {Martell}, {de Boer}, {Asplund}, {Keller}, {Sharma}, {Zucker}, {Zwitter},
  {Anguiano}, {Bacigalupo}, {Bayliss}, {Beavis}, {Bergemann}, {Campbell},
  {Cannon}, {Carollo}, {Casagrande}, {Casey}, {Da Costa}, {D'Orazi}, {Dotter},
  {Duong}, {Heger}, {Ireland}, {Kafle}, {Kos}, {Lattanzio}, {Lewis}, {Lin},
  {Lind}, {Munari}, {Nataf}, {O'Toole}, {Parker}, {Reid}, {Schlesinger},
  {Sheinis}, {Simpson}, {Stello}, {Ting}, {Traven}, {Watson}, {Wittenmyer},
  {Yong}, \& {{\v{Z}}erjal}}]{DeSilva:2015MNRAS.449.2604D}
{De Silva}, G.~M., {Freeman}, K.~C., {Bland-Hawthorn}, J., {et~al.} 2015,
  \mnras, 449, 2604, \dodoi{10.1093/mnras/stv327}

\bibitem[{{Delgado Mena} {et~al.}(2015){Delgado Mena}, {Bertr{\'a}n de Lis},
  {Adibekyan}, {Sousa}, {Figueira}, {Mortier}, {Gonz{\'a}lez Hern{\'a}ndez},
  {Tsantaki}, {Israelian}, \& {Santos}}]{Delgado_Mena:2015A&A...576A..69D}
{Delgado Mena}, E., {Bertr{\'a}n de Lis}, S., {Adibekyan}, V.~Z., {et~al.}
  2015, \aap, 576, A69, \dodoi{10.1051/0004-6361/201425433}

\bibitem[{{Demarque} {et~al.}(2004){Demarque}, {Woo}, {Kim}, \&
  {Yi}}]{Demarque:2004ApJS..155..667D}
{Demarque}, P., {Woo}, J.-H., {Kim}, Y.-C., \& {Yi}, S.~K. 2004, \apjs, 155,
  667, \dodoi{10.1086/424966}

\bibitem[{{Desidera} {et~al.}(2006){Desidera}, {Gratton}, {Lucatello}, \&
  {Claudi}}]{Desidera:2006A&A...454..581D}
{Desidera}, S., {Gratton}, R.~G., {Lucatello}, S., \& {Claudi}, R.~U. 2006,
  \aap, 454, 581, \dodoi{10.1051/0004-6361:20064896}

\bibitem[{{Desidera} {et~al.}(2004){Desidera}, {Gratton}, {Scuderi}, {Claudi},
  {Cosentino}, {Barbieri}, {Bonanno}, {Carretta}, {Endl}, {Lucatello},
  {Martinez Fiorenzano}, \& {Marzari}}]{Desidera:2004A&A...420..683D}
{Desidera}, S., {Gratton}, R.~G., {Scuderi}, S., {et~al.} 2004, \aap, 420, 683,
  \dodoi{10.1051/0004-6361:20041242}

\bibitem[{{dos Santos} {et~al.}(2016){dos Santos}, {Mel{\'e}ndez}, {do
  Nascimento}, {Bedell}, {Ram{\'\i}rez}, {Bean}, {Asplund}, {Spina},
  {Dreizler}, {Alves-Brito}, \& {Casagrande}}]{Leo:2016A&A...592A.156D}
{dos Santos}, L.~A., {Mel{\'e}ndez}, J., {do Nascimento}, J.-D., {et~al.} 2016,
  \aap, 592, A156, \dodoi{10.1051/0004-6361/201628558}

\bibitem[{{El-Badry} {et~al.}(2021){El-Badry}, {Rix}, \&
  {Heintz}}]{El-Badry:2021MNRAS.tmp..394E}
{El-Badry}, K., {Rix}, H.-W., \& {Heintz}, T.~M. 2021, \mnras,
  \dodoi{10.1093/mnras/stab323}

\bibitem[{{Epstein} {et~al.}(2010){Epstein}, {Johnson}, {Dong}, {Udalski},
  {Gould}, \& {Becker}}]{Epstein:2010ApJ...709..447E}
{Epstein}, C.~R., {Johnson}, J.~A., {Dong}, S., {et~al.} 2010, \apj, 709, 447,
  \dodoi{10.1088/0004-637X/709/1/447}

\bibitem[{{Espinoza-Rojas} {et~al.}(2021){Espinoza-Rojas}, {Chanam{\'e}},
  {Jofr{\'e}}, \& {Casamiquela}}]{Espinoza-Rojas:2021arXiv210501096E}
{Espinoza-Rojas}, F., {Chanam{\'e}}, J., {Jofr{\'e}}, P., \& {Casamiquela}, L.
  2021, arXiv e-prints, arXiv:2105.01096.
\newblock \doarXiv{2105.01096}

\bibitem[{{Gaia Collaboration} {et~al.}(2021){Gaia Collaboration}, {Brown},
  {Vallenari}, {Prusti}, {de Bruijne}, {Babusiaux}, {Biermann}, {Creevey},
  {Evans}, {Eyer}, {Hutton}, {Jansen}, {Jordi}, {Klioner}, {Lammers},
  {Lindegren}, {Luri}, {Mignard}, {Panem}, {Pourbaix}, {Randich}, {Sartoretti},
  {Soubiran}, {Walton}, {Arenou}, {Bailer-Jones}, {Bastian}, {Cropper},
  {Drimmel}, {Katz}, {Lattanzi}, {van Leeuwen}, {Bakker}, {Cacciari},
  {Casta{\~n}eda}, {De Angeli}, {Ducourant}, {Fabricius}, {Fouesneau},
  {Fr{\'e}mat}, {Guerra}, {Guerrier}, {Guiraud}, {Jean-Antoine Piccolo},
  {Masana}, {Messineo}, {Mowlavi}, {Nicolas}, {Nienartowicz}, {Pailler},
  {Panuzzo}, {Riclet}, {Roux}, {Seabroke}, {Sordo}, {Tanga}, {Th{\'e}venin},
  {Gracia-Abril}, {Portell}, {Teyssier}, {Altmann}, {Andrae}, {Bellas-Velidis},
  {Benson}, {Berthier}, {Blomme}, {Brugaletta}, {Burgess}, {Busso}, {Carry},
  {Cellino}, {Cheek}, {Clementini}, {Damerdji}, {Davidson}, {Delchambre},
  {Dell'Oro}, {Fern{\'a}ndez-Hern{\'a}ndez}, {Galluccio}, {Garc{\'\i}a-Lario},
  {Garcia-Reinaldos}, {Gonz{\'a}lez-N{\'u}{\~n}ez}, {Gosset}, {Haigron},
  {Halbwachs}, {Hambly}, {Harrison}, {Hatzidimitriou}, {Heiter},
  {Hern{\'a}ndez}, {Hestroffer}, {Hodgkin}, {Holl}, {Jan{\ss}en}, {Jevardat de
  Fombelle}, {Jordan}, {Krone-Martins}, {Lanzafame}, {L{\"o}ffler}, {Lorca},
  {Manteiga}, {Marchal}, {Marrese}, {Moitinho}, {Mora}, {Muinonen}, {Osborne},
  {Pancino}, {Pauwels}, {Petit}, {Recio-Blanco}, {Richards}, {Riello},
  {Rimoldini}, {Robin}, {Roegiers}, {Rybizki}, {Sarro}, {Siopis}, {Smith},
  {Sozzetti}, {Ulla}, {Utrilla}, {van Leeuwen}, {van Reeven}, {Abbas}, {Abreu
  Aramburu}, {Accart}, {Aerts}, {Aguado}, {Ajaj}, {Altavilla}, {{\'A}lvarez},
  {{\'A}lvarez Cid-Fuentes}, {Alves}, {Anderson}, {Anglada Varela}, {Antoja},
  {Audard}, {Baines}, {Baker}, {Balaguer-N{\'u}{\~n}ez}, {Balbinot}, {Balog},
  {Barache}, {Barbato}, {Barros}, {Barstow}, {Bartolom{\'e}}, {Bassilana},
  {Bauchet}, {Baudesson-Stella}, {Becciani}, {Bellazzini}, {Bernet}, {Bertone},
  {Bianchi}, {Blanco-Cuaresma}, {Boch}, {Bombrun}, {Bossini}, {Bouquillon},
  {Bragaglia}, {Bramante}, {Breedt}, {Bressan}, {Brouillet}, {Bucciarelli},
  {Burlacu}, {Busonero}, {Butkevich}, {Buzzi}, {Caffau}, {Cancelliere},
  {C{\'a}novas}, {Cantat-Gaudin}, {Carballo}, {Carlucci}, {Carnerero},
  {Carrasco}, {Casamiquela}, {Castellani}, {Castro-Ginard}, {Castro Sampol},
  {Chaoul}, {Charlot}, {Chemin}, {Chiavassa}, {Cioni}, {Comoretto}, {Cooper},
  {Cornez}, {Cowell}, {Crifo}, {Crosta}, {Crowley}, {Dafonte}, {Dapergolas},
  {David}, {David}, {de Laverny}, {De Luise}, {De March}, {De Ridder}, {de
  Souza}, {de Teodoro}, {de Torres}, {del Peloso}, {del Pozo}, {Delbo},
  {Delgado}, {Delgado}, {Delisle}, {Di Matteo}, {Diakite}, {Diener},
  {Distefano}, {Dolding}, {Eappachen}, {Edvardsson}, {Enke}, {Esquej}, {Fabre},
  {Fabrizio}, {Faigler}, {Fedorets}, {Fernique}, {Fienga}, {Figueras},
  {Fouron}, {Fragkoudi}, {Fraile}, {Franke}, {Gai}, {Garabato},
  {Garcia-Gutierrez}, {Garc{\'\i}a-Torres}, {Garofalo}, {Gavras}, {Gerlach},
  {Geyer}, {Giacobbe}, {Gilmore}, {Girona}, {Giuffrida}, {Gomel}, {Gomez},
  {Gonzalez-Santamaria}, {Gonz{\'a}lez-Vidal}, {Granvik},
  {Guti{\'e}rrez-S{\'a}nchez}, {Guy}, {Hauser}, {Haywood}, {Helmi}, {Hidalgo},
  {Hilger}, {H{\l}adczuk}, {Hobbs}, {Holland}, {Huckle}, {Jasniewicz},
  {Jonker}, {Juaristi Campillo}, {Julbe}, {Karbevska}, {Kervella}, {Khanna},
  {Kochoska}, {Kontizas}, {Kordopatis}, {Korn}, {Kostrzewa-Rutkowska},
  {Kruszy{\'n}ska}, {Lambert}, {Lanza}, {Lasne}, {Le Campion}, {Le Fustec},
  {Lebreton}, {Lebzelter}, {Leccia}, {Leclerc}, {Lecoeur-Taibi}, {Liao},
  {Licata}, {Lindstr{\o}m}, {Lister}, {Livanou}, {Lobel}, {Madrero Pardo},
  {Managau}, {Mann}, {Marchant}, {Marconi}, {Marcos Santos}, {Marinoni},
  {Marocco}, {Marshall}, {Martin Polo}, {Mart{\'\i}n-Fleitas}, {Masip},
  {Massari}, {Mastrobuono-Battisti}, {Mazeh}, {McMillan}, {Messina},
  {Michalik}, {Millar}, {Mints}, {Molina}, {Molinaro}, {Moln{\'a}r},
  {Montegriffo}, {Mor}, {Morbidelli}, {Morel}, {Morris}, {Mulone}, {Munoz},
  {Muraveva}, {Murphy}, {Musella}, {Noval}, {Ord{\'e}novic}, {Orr{\`u}},
  {Osinde}, {Pagani}, {Pagano}, {Palaversa}, {Palicio}, {Panahi}, {Pawlak},
  {Pe{\~n}alosa Esteller}, {Penttil{\"a}}, {Piersimoni}, {Pineau}, {Plachy},
  {Plum}, {Poggio}, {Poretti}, {Poujoulet}, {Pr{\v{s}}a}, {Pulone}, {Racero},
  {Ragaini}, {Rainer}, {Raiteri}, {Rambaux}, {Ramos}, {Ramos-Lerate}, {Re
  Fiorentin}, {Regibo}, {Reyl{\'e}}, {Ripepi}, {Riva}, {Rixon}, {Robichon},
  {Robin}, {Roelens}, {Rohrbasser}, {Romero-G{\'o}mez}, {Rowell}, {Royer},
  {Rybicki}, {Sadowski}, {Sagrist{\`a} Sell{\'e}s}, {Sahlmann}, {Salgado},
  {Salguero}, {Samaras}, {Sanchez Gimenez}, {Sanna}, {Santove{\~n}a},
  {Sarasso}, {Schultheis}, {Sciacca}, {Segol}, {Segovia}, {S{\'e}gransan},
  {Semeux}, {Shahaf}, {Siddiqui}, {Siebert}, {Siltala}, {Slezak}, {Smart},
  {Solano}, {Solitro}, {Souami}, {Souchay}, {Spagna}, {Spoto}, {Steele},
  {Steidelm{\"u}ller}, {Stephenson}, {S{\"u}veges}, {Szabados}, {Szegedi-Elek},
  {Taris}, {Tauran}, {Taylor}, {Teixeira}, {Thuillot}, {Tonello}, {Torra},
  {Torra}, {Turon}, {Unger}, {Vaillant}, {van Dillen}, {Vanel}, {Vecchiato},
  {Viala}, {Vicente}, {Voutsinas}, {Weiler}, {Wevers}, {Wyrzykowski}, {Yoldas},
  {Yvard}, {Zhao}, {Zorec}, {Zucker}, {Zurbach}, \&
  {Zwitter}}]{GaiaEDR3:2021A&A...649A...1G}
{Gaia Collaboration}, {Brown}, A.~G.~A., {Vallenari}, A., {et~al.} 2021, \aap,
  649, A1, \dodoi{10.1051/0004-6361/202039657}

\bibitem[{{Galarza} {et~al.}(2016){Galarza}, {Mel{\'e}ndez}, \&
  {Cohen}}]{Yana_Galarza:2016A&A...589A..65G}
{Galarza}, J.~Y., {Mel{\'e}ndez}, J., \& {Cohen}, J.~G. 2016, \aap, 589, A65,
  \dodoi{10.1051/0004-6361/201527477}

\bibitem[{{Gilmore} {et~al.}(2012){Gilmore}, {Randich}, {Asplund}, {Binney},
  {Bonifacio}, {Drew}, {Feltzing}, {Ferguson}, {Jeffries}, {Micela},
  {Negueruela}, {Prusti}, {Rix}, {Vallenari}, {Alfaro}, {Allende-Prieto},
  {Babusiaux}, {Bensby}, {Blomme}, {Bragaglia}, {Flaccomio}, {Fran{\c{c}}ois},
  {Irwin}, {Koposov}, {Korn}, {Lanzafame}, {Pancino}, {Paunzen},
  {Recio-Blanco}, {Sacco}, {Smiljanic}, {Van Eck}, {Walton}, {Aden}, {Aerts},
  {Affer}, {Alcala}, {Altavilla}, {Alves}, {Antoja}, {Arenou}, {Argiroffi},
  {Asensio Ramos}, {Bailer-Jones}, {Balaguer-Nunez}, {Bayo}, {Barbuy},
  {Barisevicius}, {Barrado y Navascues}, {Battistini}, {Bellas Velidis},
  {Bellazzini}, {Belokurov}, {Bergemann}, {Bertelli}, {Biazzo}, {Bienayme},
  {Bland-Hawthorn}, {Boeche}, {Bonito}, {Boudreault}, {Bouvier}, {Brandao},
  {Brown}, {de Bruijne}, {Burleigh}, {Caballero}, {Caffau}, {Calura},
  {Capuzzo-Dolcetta}, {Caramazza}, {Carraro}, {Casagrande}, {Casewell},
  {Chapman}, {Chiappini}, {Chorniy}, {Christlieb}, {Cignoni}, {Cocozza},
  {Colless}, {Collet}, {Collins}, {Correnti}, {Covino}, {Crnojevic}, {Cropper},
  {Cunha}, {Damiani}, {David}, {Delgado}, {Duffau}, {Edvardsson}, {Eldridge},
  {Enke}, {Eriksson}, {Evans}, {Eyer}, {Famaey}, {Fellhauer}, {Ferreras},
  {Figueras}, {Fiorentino}, {Flynn}, {Folha}, {Franciosini}, {Frasca},
  {Freeman}, {Fremat}, {Friel}, {Gaensicke}, {Gameiro}, {Garzon}, {Geier},
  {Geisler}, {Gerhard}, {Gibson}, {Gomboc}, {Gomez}, {Gonzalez-Fernandez},
  {Gonzalez Hernandez}, {Gosset}, {Grebel}, {Greimel}, {Groenewegen},
  {Grundahl}, {Guarcello}, {Gustafsson}, {Hadrava}, {Hatzidimitriou}, {Hambly},
  {Hammersley}, {Hansen}, {Haywood}, {Heber}, {Heiter}, {Held}, {Helmi},
  {Hensler}, {Herrero}, {Hill}, {Hodgkin}, {Huelamo}, {Huxor}, {Ibata},
  {Jackson}, {de Jong}, {Jonker}, {Jordan}, {Jordi}, {Jorissen}, {Katz},
  {Kawata}, {Keller}, {Kharchenko}, {Klement}, {Klutsch}, {Knude}, {Koch},
  {Kochukhov}, {Kontizas}, {Koubsky}, {Lallement}, {de Laverny}, {van Leeuwen},
  {Lemasle}, {Lewis}, {Lind}, {Lindstrom}, {Lobel}, {Lopez Santiago}, {Lucas},
  {Ludwig}, {Lueftinger}, {Magrini}, {Maiz Apellaniz}, {Maldonado}, {Marconi},
  {Marino}, {Martayan}, {Martinez-Valpuesta}, {Matijevic}, {McMahon},
  {Messina}, {Meyer}, {Miglio}, {Mikolaitis}, {Minchev}, {Minniti}, {Moitinho},
  {Momany}, {Monaco}, {Montalto}, {Monteiro}, {Monier}, {Montes}, {Mora},
  {Moraux}, {Morel}, {Mowlavi}, {Mucciarelli}, {Munari}, {Napiwotzki},
  {Nardetto}, {Naylor}, {Naze}, {Nelemans}, {Okamoto}, {Ortolani}, {Pace},
  {Palla}, {Palous}, {Parker}, {Penarrubia}, {Pillitteri}, {Piotto}, {Posbic},
  {Prisinzano}, {Puzeras}, {Quirrenbach}, {Ragaini}, {Read}, {Read}, {Reyle},
  {De Ridder}, {Robichon}, {Robin}, {Roeser}, {Romano}, {Royer}, {Ruchti},
  {Ruzicka}, {Ryan}, {Ryde}, {Santos}, {Sanz Forcada}, {Sarro Baro},
  {Sbordone}, {Schilbach}, {Schmeja}, {Schnurr}, {Schoenrich}, {Scholz},
  {Seabroke}, {Sharma}, {De Silva}, {Smith}, {Solano}, {Sordo}, {Soubiran},
  {Sousa}, {Spagna}, {Steffen}, {Steinmetz}, {Stelzer}, {Stempels},
  {Tabernero}, {Tautvaisiene}, {Thevenin}, {Torra}, {Tosi}, {Tolstoy}, {Turon},
  {Walker}, {Wambsganss}, {Worley}, {Venn}, {Vink}, {Wyse}, {Zaggia},
  {Zeilinger}, {Zoccali}, {Zorec}, {Zucker}, {Zwitter}, \& {Gaia-ESO Survey
  Team}}]{Gilmore:2012Msngr.147...25G}
{Gilmore}, G., {Randich}, S., {Asplund}, M., {et~al.} 2012, The Messenger, 147,
  25

\bibitem[{{Ginsburg} {et~al.}(2019){Ginsburg}, {Sip{\H{o}}cz}, {Brasseur},
  {Cowperthwaite}, {Craig}, {Deil}, {Guillochon}, {Guzman}, {Liedtke}, {Lian
  Lim}, {Lockhart}, {Mommert}, {Morris}, {Norman}, {Parikh}, {Persson},
  {Robitaille}, {Segovia}, {Singer}, {Tollerud}, {de Val-Borro}, {Valtchanov},
  {Woillez}, {Astroquery Collaboration}, \& {a subset of astropy
  Collaboration}}]{Ginsburg:2019AJ....157...98G}
{Ginsburg}, A., {Sip{\H{o}}cz}, B.~M., {Brasseur}, C.~E., {et~al.} 2019, \aj,
  157, 98, \dodoi{10.3847/1538-3881/aafc33}

\bibitem[{{Gratton} {et~al.}(2001){Gratton}, {Bonanno}, {Claudi}, {Cosentino},
  {Desidera}, {Lucatello}, \& {Scuderi}}]{Gratton:2001A&A...377..123G}
{Gratton}, R.~G., {Bonanno}, G., {Claudi}, R.~U., {et~al.} 2001, \aap, 377,
  123, \dodoi{10.1051/0004-6361:20011066}

\bibitem[{{Gustafsson}(2018{\natexlab{a}})}]{Gustafsson:2018A&A...620A..53G}
{Gustafsson}, B. 2018{\natexlab{a}}, \aap, 620, A53,
  \dodoi{10.1051/0004-6361/201833353}

\bibitem[{{Gustafsson}(2018{\natexlab{b}})}]{Gustafsson:2018A&A...616A..91G}
---. 2018{\natexlab{b}}, \aap, 616, A91, \dodoi{10.1051/0004-6361/201732354}

\bibitem[{{Hawkins} {et~al.}(2020){Hawkins}, {Lucey}, {Ting}, {Ji}, {Katzberg},
  {Thompson}, {El-Badry}, {Teske}, {Nelson}, \&
  {Carrillo}}]{Hawkins:2020MNRAS.492.1164H}
{Hawkins}, K., {Lucey}, M., {Ting}, Y.-S., {et~al.} 2020, \mnras, 492, 1164,
  \dodoi{10.1093/mnras/stz3132}

\bibitem[{Hunter(2007)}]{Hunter:4160265}
Hunter, J.~D. 2007, Computing in Science Engineering, 9, 90,
  \dodoi{10.1109/MCSE.2007.55}

\bibitem[{{Jiang} \& {Tremaine}(2010)}]{Jiang:2010MNRAS.401..977J}
{Jiang}, Y.-F., \& {Tremaine}, S. 2010, \mnras, 401, 977,
  \dodoi{10.1111/j.1365-2966.2009.15744.x}

\bibitem[{{Kamdar} {et~al.}(2019){Kamdar}, {Conroy}, {Ting}, {Bonaca},
  {Johnson}, \& {Cargile}}]{Kamdar:2019ApJ...884..173K}
{Kamdar}, H., {Conroy}, C., {Ting}, Y.-S., {et~al.} 2019, \apj, 884, 173,
  \dodoi{10.3847/1538-4357/ab44be}

\bibitem[{{Laws} \& {Gonzalez}(2001)}]{Laws:2001ApJ...553..405L}
{Laws}, C., \& {Gonzalez}, G. 2001, \apj, 553, 405, \dodoi{10.1086/320669}

\bibitem[{{Lind} {et~al.}(2009){Lind}, {Asplund}, \&
  {Barklem}}]{Lind:2009A&A...503..541L}
{Lind}, K., {Asplund}, M., \& {Barklem}, P.~S. 2009, \aap, 503, 541,
  \dodoi{10.1051/0004-6361/200912221}

\bibitem[{{Lodders}(2003)}]{Lodders:2003ApJ...591.1220L}
{Lodders}, K. 2003, \apj, 591, 1220, \dodoi{10.1086/375492}

\bibitem[{{Lorenzo-Oliveira} {et~al.}(2016){Lorenzo-Oliveira}, {Porto de
  Mello}, \& {Schiavon}}]{Diego:2016A&A...594L...3L}
{Lorenzo-Oliveira}, D., {Porto de Mello}, G.~F., \& {Schiavon}, R.~P. 2016,
  \aap, 594, L3, \dodoi{10.1051/0004-6361/201629233}

\bibitem[{{Maia} {et~al.}(2019){Maia}, {Mel{\'e}ndez}, {Lorenzo-Oliveira},
  {Spina}, \& {Jofr{\'e}}}]{Tucci:2019A&A...628A.126M}
{Maia}, M.~T., {Mel{\'e}ndez}, J., {Lorenzo-Oliveira}, D., {Spina}, L., \&
  {Jofr{\'e}}, P. 2019, \aap, 628, A126, \dodoi{10.1051/0004-6361/201935952}

\bibitem[{{Majewski} {et~al.}(2017){Majewski}, {Schiavon}, {Frinchaboy},
  {Allende Prieto}, {Barkhouser}, {Bizyaev}, {Blank}, {Brunner}, {Burton},
  {Carrera}, {Chojnowski}, {Cunha}, {Epstein}, {Fitzgerald}, {Garc{\'\i}a
  P{\'e}rez}, {Hearty}, {Henderson}, {Holtzman}, {Johnson}, {Lam}, {Lawler},
  {Maseman}, {M{\'e}sz{\'a}ros}, {Nelson}, {Nguyen}, {Nidever}, {Pinsonneault},
  {Shetrone}, {Smee}, {Smith}, {Stolberg}, {Skrutskie}, {Walker}, {Wilson},
  {Zasowski}, {Anders}, {Basu}, {Beland}, {Blanton}, {Bovy}, {Brownstein},
  {Carlberg}, {Chaplin}, {Chiappini}, {Eisenstein}, {Elsworth}, {Feuillet},
  {Fleming}, {Galbraith-Frew}, {Garc{\'\i}a}, {Garc{\'\i}a-Hern{\'a}ndez},
  {Gillespie}, {Girardi}, {Gunn}, {Hasselquist}, {Hayden}, {Hekker}, {Ivans},
  {Kinemuchi}, {Klaene}, {Mahadevan}, {Mathur}, {Mosser}, {Muna}, {Munn},
  {Nichol}, {O'Connell}, {Parejko}, {Robin}, {Rocha-Pinto}, {Schultheis},
  {Serenelli}, {Shane}, {Silva Aguirre}, {Sobeck}, {Thompson}, {Troup},
  {Weinberg}, \& {Zamora}}]{Majewski:2017AJ....154...94M}
{Majewski}, S.~R., {Schiavon}, R.~P., {Frinchaboy}, P.~M., {et~al.} 2017, \aj,
  154, 94, \dodoi{10.3847/1538-3881/aa784d}

\bibitem[{{Mart{\'\i}n} {et~al.}(2002){Mart{\'\i}n}, {Basri}, {Pavlenko}, \&
  {Lyubchik}}]{Martin:2002ApJ...579..437M}
{Mart{\'\i}n}, E.~L., {Basri}, G., {Pavlenko}, Y., \& {Lyubchik}, Y. 2002,
  \apj, 579, 437, \dodoi{10.1086/342674}

\bibitem[{McKinney(2010)}]{mckinney-proc-scipy-2010}
McKinney, W. 2010, in Proceedings of the 9th Python in Science Conference, ed.
  S.~van~der Walt \& J.~Millman, 51 -- 56

\bibitem[{{McWilliam}(1998)}]{McWilliam:1998AJ....115.1640M}
{McWilliam}, A. 1998, \aj, 115, 1640, \dodoi{10.1086/300289}

\bibitem[{{Mel{\'e}ndez} {et~al.}(2009){Mel{\'e}ndez}, {Asplund}, {Gustafsson},
  \& {Yong}}]{Melendez:2009ApJ...704L..66M}
{Mel{\'e}ndez}, J., {Asplund}, M., {Gustafsson}, B., \& {Yong}, D. 2009, \apjl,
  704, L66, \dodoi{10.1088/0004-637X/704/1/L66}

\bibitem[{{Mel{\'e}ndez} {et~al.}(2014){Mel{\'e}ndez}, {Ram{\'{\i}}rez},
  {Karakas}, {Yong}, {Monroe}, {Bedell}, {Bergemann}, {Asplund}, {Tucci Maia},
  {Bean}, {do Nascimento}, {Bazot}, {Alves-Brito}, {Freitas}, \&
  {Castro}}]{Melendez:2014ApJ...791...14M}
{Mel{\'e}ndez}, J., {Ram{\'{\i}}rez}, I., {Karakas}, A.~I., {et~al.} 2014,
  \apj, 791, 14, \dodoi{10.1088/0004-637X/791/1/14}

\bibitem[{{Mel{\'e}ndez} {et~al.}(2017){Mel{\'e}ndez}, {Bedell}, {Bean},
  {Ram{\'\i}rez}, {Asplund}, {Dreizler}, {Yan}, {Shi}, {Lind}, {Ferraz-Mello},
  {Galarza}, {dos Santos}, {Spina}, {Maia}, {Alves-Brito}, {Monroe}, \&
  {Casagrande}}]{Melendez:2017A&A...597A..34M}
{Mel{\'e}ndez}, J., {Bedell}, M., {Bean}, J.~L., {et~al.} 2017, \aap, 597, A34,
  \dodoi{10.1051/0004-6361/201527775}

\bibitem[{{Nagar} {et~al.}(2020){Nagar}, {Spina}, \&
  {Karakas}}]{Nagar:2020ApJ...888L...9N}
{Nagar}, T., {Spina}, L., \& {Karakas}, A.~I. 2020, \apjl, 888, L9,
  \dodoi{10.3847/2041-8213/ab5dc6}

\bibitem[{{Nelson} {et~al.}(2021){Nelson}, {Ting}, {Hawkins}, {Ji}, {Kamdar},
  \& {El-Badry}}]{Nelson:2021arXiv210412883N}
{Nelson}, T., {Ting}, Y.-S., {Hawkins}, K., {et~al.} 2021, arXiv e-prints,
  arXiv:2104.12883.
\newblock \doarXiv{2104.12883}

\bibitem[{{Ness} {et~al.}(2018){Ness}, {Rix}, {Hogg}, {Casey}, {Holtzman},
  {Fouesneau}, {Zasowski}, {Geisler}, {Shetrone}, {Minniti}, {Frinchaboy}, \&
  {Roman-Lopes}}]{Ness:2018ApJ...853..198N}
{Ness}, M., {Rix}, H.~W., {Hogg}, D.~W., {et~al.} 2018, \apj, 853, 198,
  \dodoi{10.3847/1538-4357/aa9d8e}

\bibitem[{{Noyes} {et~al.}(1984){Noyes}, {Hartmann}, {Baliunas}, {Duncan}, \&
  {Vaughan}}]{Noyes:1984ApJ...279..763N}
{Noyes}, R.~W., {Hartmann}, L.~W., {Baliunas}, S.~L., {Duncan}, D.~K., \&
  {Vaughan}, A.~H. 1984, \apj, 279, 763, \dodoi{10.1086/161945}

\bibitem[{{Oetjens} {et~al.}(2020){Oetjens}, {Carone}, {Bergemann}, \&
  {Serenelli}}]{Oetjens:2020A&A...643A..34O}
{Oetjens}, A., {Carone}, L., {Bergemann}, M., \& {Serenelli}, A. 2020, \aap,
  643, A34, \dodoi{10.1051/0004-6361/202038653}

\bibitem[{{Oh} {et~al.}(2018){Oh}, {Price-Whelan}, {Brewer}, {Hogg}, {Spergel},
  \& {Myles}}]{Oh:2018ApJ...854..138O}
{Oh}, S., {Price-Whelan}, A.~M., {Brewer}, J.~M., {et~al.} 2018, \apj, 854,
  138, \dodoi{10.3847/1538-4357/aaab4d}

\bibitem[{{Pinsonneault} {et~al.}(2001){Pinsonneault}, {DePoy}, \&
  {Coffee}}]{Pinsonneault:2001ApJ...556L..59P}
{Pinsonneault}, M.~H., {DePoy}, D.~L., \& {Coffee}, M. 2001, \apjl, 556, L59,
  \dodoi{10.1086/323531}

\bibitem[{Price-Whelan(2017)}]{gala}
Price-Whelan, A.~M. 2017, The Journal of Open Source Software, 2,
  \dodoi{10.21105/joss.00388}

\bibitem[{{Ram{\'\i}rez} {et~al.}(2007){Ram{\'\i}rez}, {Allende Prieto}, \&
  {Lambert}}]{Ramirez:2007A&A...465..271R}
{Ram{\'\i}rez}, I., {Allende Prieto}, C., \& {Lambert}, D.~L. 2007, \aap, 465,
  271, \dodoi{10.1051/0004-6361:20066619}

\bibitem[{{Ram{\'\i}rez} {et~al.}(2012){Ram{\'\i}rez}, {Fish}, {Lambert}, \&
  {Allende Prieto}}]{Ramirez:2012ApJ...756...46R}
{Ram{\'\i}rez}, I., {Fish}, J.~R., {Lambert}, D.~L., \& {Allende Prieto}, C.
  2012, \apj, 756, 46, \dodoi{10.1088/0004-637X/756/1/46}

\bibitem[{{Ram{\'\i}rez} {et~al.}(2019){Ram{\'\i}rez}, {Khanal}, {Lichon},
  {Chanam{\'e}}, {Endl}, {Mel{\'e}ndez}, \&
  {Lambert}}]{Ramirez:2019MNRAS.490.2448R}
{Ram{\'\i}rez}, I., {Khanal}, S., {Lichon}, S.~J., {et~al.} 2019, \mnras, 490,
  2448, \dodoi{10.1093/mnras/stz2709}

\bibitem[{{Ram{\'\i}rez} {et~al.}(2011){Ram{\'\i}rez}, {Mel{\'e}ndez},
  {Cornejo}, {Roederer}, \& {Fish}}]{Ramirez:2011ApJ...740...76R}
{Ram{\'\i}rez}, I., {Mel{\'e}ndez}, J., {Cornejo}, D., {Roederer}, I.~U., \&
  {Fish}, J.~R. 2011, \apj, 740, 76, \dodoi{10.1088/0004-637X/740/2/76}

\bibitem[{{Ram{\'\i}rez} {et~al.}(2014){Ram{\'\i}rez}, {Mel{\'e}ndez}, {Bean},
  {Asplund}, {Bedell}, {Monroe}, {Casagrande}, {Schirbel}, {Dreizler}, {Teske},
  {Tucci Maia}, {Alves-Brito}, \& {Baumann}}]{Ramirez:2014A&A...572A..48R}
{Ram{\'\i}rez}, I., {Mel{\'e}ndez}, J., {Bean}, J., {et~al.} 2014, \aap, 572,
  A48, \dodoi{10.1051/0004-6361/201424244}

\bibitem[{{Ram{\'\i}rez} {et~al.}(2015){Ram{\'\i}rez}, {Khanal}, {Aleo},
  {Sobotka}, {Liu}, {Casagrande}, {Mel{\'e}ndez}, {Yong}, {Lambert}, \&
  {Asplund}}]{Ramirez:2015ApJ...808...13R}
{Ram{\'\i}rez}, I., {Khanal}, S., {Aleo}, P., {et~al.} 2015, \apj, 808, 13,
  \dodoi{10.1088/0004-637X/808/1/13}

\bibitem[{{Saffe} {et~al.}(2015){Saffe}, {Flores}, \&
  {Buccino}}]{Saffe:2015A&A...582A..17S}
{Saffe}, C., {Flores}, M., \& {Buccino}, A. 2015, \aap, 582, A17,
  \dodoi{10.1051/0004-6361/201526644}

\bibitem[{{Saffe} {et~al.}(2017){Saffe}, {Jofr{\'e}}, {Martioli}, {Flores},
  {Petrucci}, \& {Jaque Arancibia}}]{Saffe:2017A&A...604L...4S}
{Saffe}, C., {Jofr{\'e}}, E., {Martioli}, E., {et~al.} 2017, \aap, 604, L4,
  \dodoi{10.1051/0004-6361/201731430}

\bibitem[{{Sneden}(1973)}]{Sneden:1973PhDT.......180S}
{Sneden}, C.~A. 1973, PhD thesis, THE UNIVERSITY OF TEXAS AT AUSTIN.

\bibitem[{{Stassun} \& {Torres}(2021)}]{Stassun:2021ApJ...907L..33S}
{Stassun}, K.~G., \& {Torres}, G. 2021, \apjl, 907, L33,
  \dodoi{10.3847/2041-8213/abdaad}

\bibitem[{{Terlouw} \& {Vogelaar}(2015)}]{KapteynPackage}
{Terlouw}, J.~P., \& {Vogelaar}, M.~G.~R. 2015, {Kapteyn Package, version 2.3},
  {Kapteyn Astronomical Institute}, Groningen

\bibitem[{{Teske} {et~al.}(2016{\natexlab{a}}){Teske}, {Khanal}, \&
  {Ram{\'\i}rez}}]{Teske:2016ApJ...819...19T}
{Teske}, J.~K., {Khanal}, S., \& {Ram{\'\i}rez}, I. 2016{\natexlab{a}}, \apj,
  819, 19, \dodoi{10.3847/0004-637X/819/1/19}

\bibitem[{{Teske} {et~al.}(2016{\natexlab{b}}){Teske}, {Shectman}, {Vogt},
  {D{\'\i}az}, {Butler}, {Crane}, {Thompson}, \&
  {Arriagada}}]{Teske:2016AJ....152..167T}
{Teske}, J.~K., {Shectman}, S.~A., {Vogt}, S.~S., {et~al.} 2016{\natexlab{b}},
  \aj, 152, 167, \dodoi{10.3847/0004-6256/152/6/167}

\bibitem[{{Ting} {et~al.}(2012){Ting}, {De Silva}, {Freeman}, \&
  {Parker}}]{Ting:2012MNRAS.427..882T}
{Ting}, Y.-S., {De Silva}, G.~M., {Freeman}, K.~C., \& {Parker}, S.~J. 2012,
  \mnras, 427, 882, \dodoi{10.1111/j.1365-2966.2012.22028.x}

\bibitem[{{Tody}(1986)}]{Tody:1986SPIE..627..733T}
{Tody}, D. 1986, in \procspie, Vol. 627, Instrumentation in astronomy VI, ed.
  D.~L. {Crawford}, 733, \dodoi{10.1117/12.968154}

\bibitem[{{Tokovinin}(2014)}]{Tokovinin:2014AJ....147...87T}
{Tokovinin}, A. 2014, \aj, 147, 87, \dodoi{10.1088/0004-6256/147/4/87}

\bibitem[{{Tull} {et~al.}(1995){Tull}, {MacQueen}, {Sneden}, \&
  {Lambert}}]{Tull:1995PASP..107..251T}
{Tull}, R.~G., {MacQueen}, P.~J., {Sneden}, C., \& {Lambert}, D.~L. 1995,
  \pasp, 107, 251, \dodoi{10.1086/133548}

\bibitem[{{van der Walt} {et~al.}(2011){van der Walt}, {Colbert}, \&
  {Varoquaux}}]{van_der_Walt:2011CSE....13b..22V}
{van der Walt}, S., {Colbert}, S.~C., \& {Varoquaux}, G. 2011, Computing in
  Science and Engineering, 13, 22, \dodoi{10.1109/MCSE.2011.37}

\bibitem[{{Wasson} \& {Kallemeyn}(1988)}]{Wasson:1988RSPTA.325..535W}
{Wasson}, J.~T., \& {Kallemeyn}, G.~W. 1988, Philosophical Transactions of the
  Royal Society of London Series A, 325, 535, \dodoi{10.1098/rsta.1988.0066}

\bibitem[{{Wright} {et~al.}(2004){Wright}, {Marcy}, {Butler}, \&
  {Vogt}}]{Wright:2004ApJS..152..261W}
{Wright}, J.~T., {Marcy}, G.~W., {Butler}, R.~P., \& {Vogt}, S.~S. 2004, \apjs,
  152, 261, \dodoi{10.1086/386283}

\bibitem[{{Yana Galarza} {et~al.}(2016){Yana Galarza}, {Mel{\'e}ndez},
  {Ram{\'{\i}}rez}, {Yong}, {Karakas}, {Asplund}, \&
  {Liu}}]{Yana_Galarza:2016A&A...589A..17Y}
{Yana Galarza}, J., {Mel{\'e}ndez}, J., {Ram{\'{\i}}rez}, I., {et~al.} 2016,
  \aap, 589, A17, \dodoi{10.1051/0004-6361/201527912}

\bibitem[{{Yana Galarza} {et~al.}(2021){Yana Galarza}, {L{\'o}pez-Valdivia},
  {Lorenzo-Oliveira}, {Reggiani}, {Mel{\'e}ndez}, {Gamarra-S{\'a}nchez},
  {Flores}, {Portal-Rivera}, {Miquelarena}, {Ponte}, {Schlaufman}, \& {Vargas
  Auccalla}}]{Yana-Galarza:2021MNRAS.504.1873Y}
{Yana Galarza}, J., {L{\'o}pez-Valdivia}, R., {Lorenzo-Oliveira}, D., {et~al.}
  2021, \mnras, 504, 1873, \dodoi{10.1093/mnras/stab987}

\bibitem[{{Yi} {et~al.}(2001){Yi}, {Demarque}, {Kim}, {Lee}, {Ree}, {Lejeune},
  \& {Barnes}}]{Yi:2001ApJS..136..417Y}
{Yi}, S., {Demarque}, P., {Kim}, Y.-C., {et~al.} 2001, \apjs, 136, 417,
  \dodoi{10.1086/321795}

\end{thebibliography}
\bibliographystyle{aasjournal}



\end{document}